\documentclass[prc,aps,preprint,showkeys,lengthcheck,twocolumn,nofootinbib,notitlepage,floatfix,superscriptaddress, nolongbibliography,amsmath,amssymb]{revtex4-2}

\usepackage{graphicx}
\usepackage{dcolumn}
\usepackage{bm}

\usepackage{xcolor}

\begin{document}

\preprint{APS/123-QED}

\title{
Toward a Direct Measurement of Partial Restoration of Chiral Symmetry\\at J-PARC E16 via Density-induced Chiral Mixing
}

\author{Ren~Ejima}
 \email{ejima@quark.hiroshima-u.ac.jp}
\affiliation{%
 Graduate School of Advanced Science and Engineering,
 Hiroshima University, 1-3-1 Kagamiyama, Higashi-Hiroshima, Hiroshima 739-8526, Japan}
 \affiliation{
 International Institute for Sustainability with Knotted Chiral Meta Matter (WPI-SKCM$^2$), Hiroshima University, 1-3-1 Kagamiyama, Higashi-Hiroshima, Hiroshima 739-8526, Japan
}%

\author{Philipp~Gubler}
 \email{gubler@post.j-parc.jp}
\affiliation{%
 Advanced Science Research Center, Japan Atomic Energy Agency,
 2-4 Shirakata Shirane, Tokai-mura, Naka-gun, Ibaraki 319-1195, Japan
}%

\author{Chihiro~Sasaki}
 \email{chihiro.sasaki@uwr.edu.pl}
\affiliation{%
 Institute of Theoretical Physics, University of Wroclaw, PL-50204 Wroclaw, Poland}
 \affiliation{
 International Institute for Sustainability with Knotted Chiral Meta Matter (WPI-SKCM$^2$), Hiroshima University, 1-3-1 Kagamiyama, Higashi-Hiroshima, Hiroshima 739-8526, Japan
}%

\author{Kenta~Shigaki}
 \email{shigaki@hiroshima-u.ac.jp}
\affiliation{%
 Graduate School of Advanced Science and Engineering,
 Hiroshima University, 1-3-1 Kagamiyama, Higashi-Hiroshima, Hiroshima 739-8526, Japan}
 \affiliation{
 International Institute for Sustainability with Knotted Chiral Meta Matter (WPI-SKCM$^2$), Hiroshima University, 1-3-1 Kagamiyama, Higashi-Hiroshima, Hiroshima 739-8526, Japan
}%

\date{\today}

\begin{abstract}
The degeneracy of chiral partners is an ideal signal for measuring the restoration of the spontaneously broken chiral symmetry in QCD. 
In this work, we investigate the observability of the $\phi$ - $f_1(1420)$ degeneracy in the J-PARC E16 experiment, which measures di-electrons emitted from 30 GeV pA collisions.
We for this purpose make use of an effective Lagrangian approach, which naturally incorporates the broken charge-conjugation symmetry in nuclear matter and the ensuing anomaly-induced mixing between vector and axial-vector mesons, to compute the spectral function
relevant for the experimental measurement. The real-time dynamics of the pA collision is obtained from a transport simulation. 
Including experimental background and resolution effects on top of that, we find that a signal of the $\phi$ - $f_1(1420)$ mixing can be observed around 2.5 $\sigma$ with the Run2 statistics planned for the J-PARC E16 experiment with an ideal mixing strength.
\end{abstract}

\maketitle


\section{\label{intro}Introduction}
Spontaneous breaking of the QCD chiral symmetry plays an essential role in the emergence of hadron masses.
It is expected on theoretical grounds \cite{Hayano_Hatsuda,PhysRevC.63.054907,Rapp:2009yu,FUKUSHIMA201399}, that the spontaneously broken chiral symmetry is restored in a medium with finite temperature or density, such as that produced in high-energy heavy-ion collisions or fixed-target experiments. 
However, the corresponding experimental verification has so far hardly been successful. Multiple experiments have attempted to measure the in-medium masses of hadrons, most commonly the vector meson masses via di-leptons \cite{Meninno_2020, URAS2014218,NA60:2008ctj, Rapp_2013}, as leptons do not interact strongly with the medium and hence provide a relatively clean signal.

Among other experimental efforts, the KEK-PS E325 experiment measured the mass distribution of vector mesons in nuclei, produced from 12 GeV pC and pCu collisions, and found a significant modification from the expected vacuum shape \cite{E325_Muto}. 
Specifically, the di-lepton data for the Cu target case were reported to be consistent with a $\phi$ meson mass shift of 3\%, when sufficiently slow $\phi$ mesons were selected. On the other hand, at the JLab CLAS experiment, no similar mass shift was observed \cite{CLAS:2007dll}. As for the $\rho$ and $\omega$ mesons, a significant mass shift was observed in the E325 experiment, whereas again no such signal was seen at CLAS \cite{CLAS:2007dll}. Similarly, an attempt was made to measure the mass modification of the $\omega$ meson in the CBELSA-TAPS experiment, but no significant change was observed either \cite{taps}. 
To improve this unclear situation with many contradictory results, the J-PARC E16 experiment will make use of the world's most intense proton beam to achieve large statistics and of spectrometers specialized for low mass di-electron measurements~\cite{Ichikawa_QM}. 
The goal of the experiment will be a systematic study of nuclear matter effects by using various targets to change the size of the medium.

However, even if the the J-PARC E16 or any other experiment shows a change in the mass distribution, it is not immediately possible to infer that the chiral symmetry has been restored and the hadron mass has changed.
The deformation must take into account not only the change on the vector meson due to the restoration of chiral symmetry, but also other interaction effects between the medium and the vector meson (such interactions may or may not be related to chiral symmetry). 
While strong modifications observed so far in dilepton production are an apparent signature of such non-trivial interactions, it remains elusive if they are traced back directly to the meson mass~\cite{CERES:1995vll,Rapp:1999ej,NA60:2006ymb,vanHees:2006ng}.
Indeed, many hadronic models have been used to quantify the spectral function of vector mesons at finite temperature and/or density, and most of them concluded that the interactions of the vector mesons with the medium cause both width broadening and a modified mass \cite{FRIMAN1997496,MUEHLICH2006187,Oset:2012ap,unknown,Lenske:2023mis}. 

Unlike hadronic models, the QCD sum rule approach directly relates certain integrals of hadronic spectral functions with expectation values of QCD operators (some of which are order parameters of chiral symmetry) both in vacuum and in the medium. 
According to recent QCD sum rule studies, the $\phi$ meson is expected to have a mass shift that strongly depends on the strangeness sigma term $\sigma_{sN} = m_s \langle N|\bar{s}s|N\rangle$, which governs the restoration of chiral symmetry in the strange quark sector at linear order in density.  
Specifically, the $\phi$ meson was predicted to have a small but positive mass shift for $\sigma_{sN}$ values smaller than 35 MeV, which turns into a negative one as $\sigma_{sN}$ increases \cite{PhysRevD.90.094002}. 
Combining this result with the latest lattice QCD calculations of $\sigma_{sN}$, which point to a value of 52.9(7.0) MeV \cite{FlavourLatticeAveragingGroupFLAG:2021npn}, one obtains a small negative mass shift of about 5 MeV at normal nuclear matter density.

It is worth mentioning here that recent measurements of the $\phi N$ channel correlation function by the ALICE collaboration have established the existence of an attractive interaction between the $\phi$ meson and protons \cite{PhysRevLett.127.172301}. 
This correlation function is in qualitative agreement with the lattice QCD calculation of the HAL QCD collaboration \cite{Chizzali:2022pjd}, even though it should be noted that the ALICE measurement contains both spin 1/2 and 3/2 components, while the HAL QCD result is only done for spin 3/2 and that the two findings can thus not be directly compared. 
The relatively large scattering length reported by the ALICE and HAL QCD collaborations seem to be in contradiction with most hadronic models and the QCD sum rule results (see, for example, the discussion in the introduction of Ref.~\cite{Gubler:2024ovg}), hence further work is clearly needed to reconcile the different theoretical and experimental results (some first steps in this direction can be found in Refs.~\cite{Feijoo:2024bvn,Abreu:2024qqo}).

To overcome the difficulty in interpreting the mass distribution due to the interaction between the medium and hadrons, we in this work propose to observe the degeneracy of the chiral partner via a phenomenon that mixes a vector with an axial-vector states in a medium, known as chiral mixing. Applying a chiral transformation to a hadron field will rotate it among the members of its multiplet and at the same time change its parity. All members of the chiral multiplet with opposite parity, that the original hadron can be rotated into, are called chiral partners. 
Since chiral symmetry is spontaneously broken in vacuum, the chiral partners turn out to have different masses.
However, at finite baryon density, the broken chiral symmetry is partially restored, and the mass distributions of the chiral partners are expected to approach each other.
Such degeneracy of the chiral partners cannot be explained by interactions with the medium that are not related to chiral symmetry, but constitutes direct evidence for its restoration. 

The chiral partners of vector mesons are axial-vector mesons, which cannot decay directly into di-leptons, but only through the process of chiral mixing. 
Specifically, in this work, we assume the chiral partner of the $\phi$ meson to be the $f_1(1420)$~\footnote{
The $\phi$ meson and $f_1(1420)$ meson are the admixtures of the flavor-$SU(3)$ octet and singlet components, and they become the exact chiral partners only when the disconnected diagrams are neglected~\cite{PhysRevD.105.014014,lee2024chiralsymmetrybreakingchiral}.
The contribution of those diagrams is known to be very small in Lattice QCD simulations at finite temperature~\cite{PhysRevD.100.094510}. We assume that this violation remains small in cold and dense QCD.
}
which so far has not been observed in any di-lepton measurement. 
One reason for this is that many heavy-ion collision experiments are conducted at rather high temperatures, 
for which the chiral mixing nearly vanishes at the chiral crossover~\cite{SASAKI2009350c}.
A footprint of the $f_1(1420)$ can remain in the in-medium vector spectral function, but its signature in thermal dilepton production is to a large extent Boltzmann-suppressed.

The situation is totally different in a cold and dense environment. 
There in fact exists a new class of chiral mixing induced by baryon density via chiral anomaly, leading to a direct mixing between the vector and axial-vector mesons at leading order~\cite{PhysRevLett.99.141602}.
Unlike the conventional mixing induced by pions, this density-induced chiral mixing never vanishes but becomes maximal when the chiral symmetry is restored~\cite{SASAKI2020135172}.

Due to the nature of the measured pA collisions, the J-PARC E16 experiment is conducted at zero temperature, which means that the thermal suppression of the $f_1(1420)$ does not play any significant role. 
In addition, as will be discussed in the next section, this mixing modifies the dispersion relation, and is enhanced at both higher momentum and with closer degeneracy of the chiral partners. 
At the same time, however, particles with a larger momentum in the lab frame will leave the target nucleus faster, which suppresses the finite density effects that can be measured in pA collisions. 
Therefore, as will be seen in Section\,\ref{result} in more detail, it is crucial to select an appropriate momentum region, where the chiral mixing is strong enough, but at the same time the relevant particles have enough time to interact with the medium. 

The most often discussed channel of chiral mixing is $\rho$ - $a_1$, which is however known to have a rather complicated spectrum due to the significant effects of nucleon resonances \cite{PhysRevD.106.054034}. 
Therefore, we here focus on the $\phi$ - $f_1(1420)$ mixing, for which the peaks in the spectrum are relatively well isolated from the surrounding resonances.

In this paper, we calculate the mass distribution expected at the J-PARC E16 experiment considering chiral mixing at finite density and discuss whether the mixing and/or degeneracy of the chiral partner is observable under realistic experimental conditions.

The paper is organized as follows: In Section \ref{theory}, we briefly introduce an effective Lagrangian with chiral mixing at finite density and its characteristics. 
In Section \ref{estimation}, we show how to calculate the expected di-electron spectrum at the J-PARC E16 experiment, based on our effective Lagrangian approach and realistic estimates of experimental backgound and resolution effects. 
In Section \ref{result}, we discuss the results and conclude the paper in Section \ref{summary} with a summary.

\section{\label{theory}Chiral Mixing in Dense Matter}

Chiral mixing is expected to play a key role in probing in-medium modifications of vector mesons leading to the restoration of chiral symmetry.
We briefly introduce a novel class of chiral mixing induced directly by finite baryon density in cold dense matter, based on Refs.~\cite{SASAKI2020135172,PhysRevD.106.054034}.

The standard interaction Lagrangian for the nucleon and omega meson reads
\begin{equation}
    \mathcal{L}_\omega 
    = g_\omega\,\bar{N}\gamma^\mu\omega_\mu N\,.
\end{equation}
Its time-component, proportional to the density operator $N^\dagger N$, gives rise to the Fermi momentum $p_F = \sqrt{(\mu_B - g_\omega\omega_0)^2 - m_N^2}$ with the nucleon mass $m_N$, whereas this operator explicitly breaks charge-conjugation invariance.
Consequently, the Lagrangian includes a set of new operators that preserve only parity and chiral invariance.
The lowest-dimensional term at leading order reads
\begin{equation}
\mathcal{L}_{\rm{mix}}=2c\epsilon^{0\mu\nu\lambda}\rm{tr}\left[\partial_\mu V_\nu\cdot A_\lambda+\partial_\mu A_\nu\cdot V_\lambda\right]\,,
\label{eq:mix}
\end{equation}
where $V_\mu$ represents the vector meson field, $A_\mu$ the axial-vector meson field, and $c$ a mixing strength~\footnote{
We will use the symbol $c$ to represent exclusively the mixing strength, not the speed of light, throughout this paper. 
}.
This term describes a direct mixing between the vector and axial-vector fields at tree level. Thus, at finite density, chiral mixing emerges in a totally distinct mechanism from the mixing due to pion loops at finite temperature according to the celebrated Dey-Eletsky-Ioffe mixing theorem~\cite{Dey:1990ba}. 
The emergence of the density-induced mixing can be traced back to the Chern-Simons term in a 5-dimensional theory as proposed within a holographic QCD approach at finite $\mu_B$~\cite{PhysRevLett.99.141602}. The same is alternatively deduced~\cite{Harada:2009cn} from the Wess-Zumino-Witten (WZW) action in 4 dimensions~\cite{KAISER1990671}.

Whereas the existence of the above operator is fully expected based on symmetry arguments, its relevance in physics entirely relies on the mixing strength, the parameter $c$.
The holographic model predicts a rather strong mixing, $c \simeq 1$ GeV at the saturation density $\rho_0$~\cite{PhysRevLett.99.141602}. Such a drastic mixing results in the vector-meson condensation at $\rho = 1.1\,\rho_0$, which is most likely an artifact of the large $N_c$ assumption in the theoretical approach with holography. 
A more realistic number is found in the standard 4-dim approach in the mean-field approximation, namely $c \simeq 0.1$ GeV at $\rho_0$~\cite{Harada:2009cn}.
Given the fact that its QCD-proper value is not known at finite density, we will treat the mixing $c$ as a parameter in the following calculations.

\begin{figure}[t!]
\includegraphics[width=85mm]{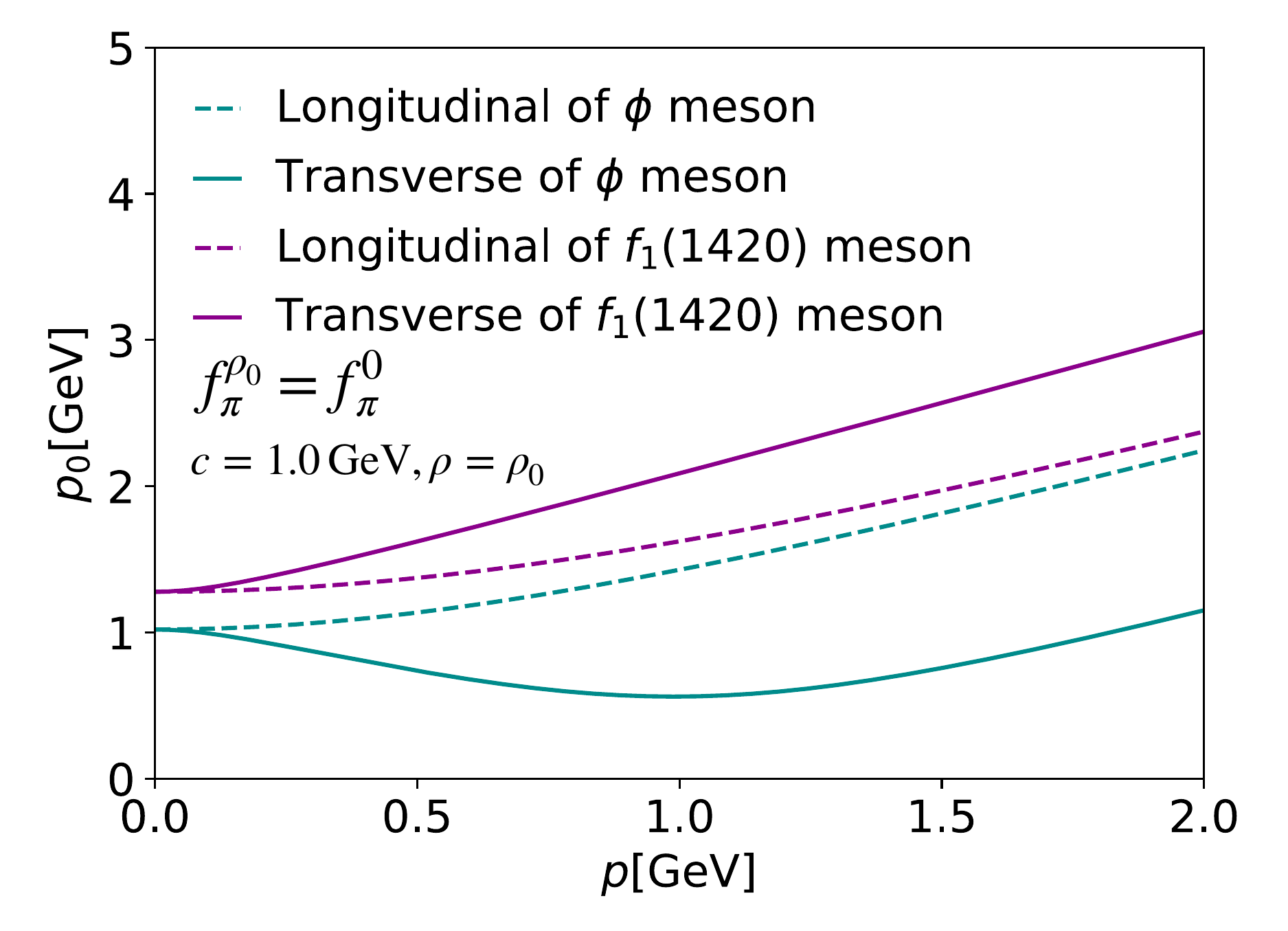}
\caption{\label{dispersion} 
The dispersion relations of $\phi$ and $f_1(1420)$ mesons with the mixing strength of $c=1$ GeV.
The masses at rest are $m_V=1.02\,{\rm{GeV}}$ and $m_A=1.42\,{\rm{GeV}}$~\cite{ParticleDataGroup:2024cfk}.
}
\end{figure}

Equation (\ref{eq:mix}) leads to the following dispersion relations for the transverse modes of vector states:
\begin{align}
    {p_0}^2-\vec{p}^2=\frac{1}{2}\left[ m_V^2+m_A^2\pm\sqrt{(m_A^2-m_V^2)^2+16c^2\vec{p}^2} \right]\,,
    \label{eq:dispersion_rel}
\end{align}
with the lower sign for the vector and the upper one for the axial-vector mesons, respectively.
Note that the longitudinal modes obey the standard dispersion relations.
The dispersion relations of the $\phi$ and $f_1$ states are shown in Fig.~\ref{dispersion}, where we take the mixing parameter of $c=1$ GeV for illustrative purposes only.

To recall the importance of chiral mixing near the chiral restoration, we expand the above dispersion relations for a small 3-momentum $\vec{p}$ to obtain
\begin{equation}
    p_0^2\simeq m_{A,V}^2+\left( 1\pm\frac{4c^2}{m_A^2-m_V^2} \right)\vec{p}^2\,.
    \label{eq:modified_dispersion_expanded}
\end{equation}
One readily finds that the effect of chiral mixing is enhanced as the masses of the vector meson and axial-vector meson become degenerate.
Note, however, that a divergence in Eq.\,(\ref{eq:modified_dispersion_expanded}) when the masses are completely degenerate is just superficial, since it is only valid for $\delta m^2 = m_A^2-m_V^2 \gg 16c^2 \vec{p}^2$.
In fact, the original dispersion laws in Eq.~(\ref{eq:dispersion_rel}) remain finite in the limit of $\delta m^2 \to 0$.

\section{\label{estimation}Construction of Invariant Mass Spectrum at J-PARC E16 Experiment}
We will in this section discuss the magnitude of the chiral mixing effect for the di-electron spectrum to be measured at the J-PARC E16 experiment, where $\phi$ mesons are produced in the target nucleus by injecting a proton beam. These $\phi$ mesons, together with all other modes of the same quantum number are observed by reconstructing the invariant mass of the di-electrons. 
We here construct this invariant mass distribution implementing chiral mixing at finite density as 
\begin{align}
\frac{dN_\phi}{ds}=\frac{\alpha^2}{\pi^3 s}\int{\rm{Im}}G_V(\vec{p},s,\rho)\frac{dN}{d\vec{p}d\rho dt}\frac{d\vec{p}}{2p_0}d\rho dt,
\label{eq:InvMass}
\end{align}
\begin{align}
\frac{dN}{dm}=\int \left[\frac{dN_\phi}{ds}+\frac{dN_{\rm{Bkg}}}{ds}\right]g(m-s)ds.
\label{eq:resolution_conv}
\end{align}
Here, the spectral function ${\rm{Im}}G_V$ depends on momentum, energy, and density, taking into account effects of chiral mixing at finite density. $dN/dpd\rho dt$ is the density and momentum distribution felt by the $\phi$ meson at the time and point of its decay, which is used as an appropriate weight in the spectral function integral. 
This weight includes the probability of the $\phi$ meson to decay into a dilepton pair. 
The invariant mass distribution can be defined as a convolution of $\frac{dN_\phi}{ds}$ plus the background $\frac{dN_{\rm{Bkg}}}{ds}$ by the mass resolution of the J-PARC E16 spectrometer, $g(m)$ for which we use a Gaussian.

A more detailed description of each component follows.

\subsection{In-medium spectral function}
We begin with the longitudinal and transverse parts of the current-current correlation function in the presence of chiral mixing at finite density~\cite{Harada:2009cn}:
\begin{equation}
G_V^L
=
\left(\frac{g_V}{m_V}\right)^2\frac{-s}{D_V^L}\,,
\quad
G_V^T
=
\left(\frac{g_V}{m_V}\right)^2
\frac{-sD_A^T + 4c^2\vec{p}^2}{D_V^T D_A^T - 4c^2\vec{p}^2}\,,
\end{equation}
with $s=p_0^2 - \vec{p}^2$, the coupling of vector meson to the vector current $g_V$ and the propagator inverse without chiral mixing $D_{V,A}^{L,T} = s - m_{V,A}^2 - \Sigma_{V,A}^{L,T}(s)$. The self-energy $\Sigma_V^{L,T}$ is computed with the modified kaons in nuclear matter, which will be outlined below based on Ref.~\cite{PhysRevD.106.054034}.
The spin-averaged correlator reads
\begin{equation}
G_V = \frac{1}{3}\left( G_V^L + 2G_V^T \right)\,,
\end{equation}
and its imaginary part defines the vector spectral function.
In the following, the labels $(V,A)$ refer exclusively to the iso-singlet $(\phi,f_1(1420))$ states. 

\subsubsection{Kaons in nuclear matter}
The $\phi$ mesons mainly decay into $K\bar{K}$, and the dressing of kaon cloud is the primary origin of modifications of $\phi$ mesons in nuclear matter.
The masses of the kaon and anti-kaon in the mean field approximation are expressed as \cite{Kaplan:1987sc}
\begin{align}
m^*_K=&\left[ m_K^2-a_K\rho_S+(b_K\rho)^2 \right]^{1/2}+b_K\rho\\
m^*_{\bar{K}}=&\left[ m_K^2-a_{\bar{K}}\rho_S+(b_K\rho)^2 \right]^{1/2}-b_K\rho
\end{align}
where $\rho_S$ is the scalar density and the three parameters are $a_K=a_{\bar{K}}=\Sigma_{KN}/f_\pi^2$ and $b_K=3/(8f_\pi^2)$.
Since the kaon-nucleon sigma term suffers from large uncertainties, we instead use $a_K=0.22$ GeV and $a_{\bar{K}} =0.45$ GeV given in \cite{Li1997KaonPI} where those parameters were determined to reproduce the data of kaon production in heavy-ion experiments.
Given those parameters, one finds $m_K^*=510~{\rm{MeV}}$ and $m_{\bar{K}}^*=380~{\rm{MeV}}$ at the normal nuclear density.

With the modified kaon masses, we obtain the width of the $\phi$ meson in the medium as
\begin{align}
    &\Gamma_{\phi}^*(s)=\frac{g_{\phi K\bar{K}}^2}{3\pi}\frac{k(s)^3}{s}\,,\\
    k(s)=\frac{1}{2\sqrt{s}}[ (s-&(m^*_K+m^*_{\bar{K}})^2)(s-(m^*_K-m^*_{\bar{K}})^2) ]^{1/2}\,,
\end{align}
with the coupling constant being $g_{\phi K\bar{K}}/4\pi=1.69$~\cite{CHUNG1998357}. 
A systematic study in the SU(3) coupled channel approach has shown that the $\phi$ has a tiny mass shift in nuclear matter~\cite{OSET2001616}. Therefore, for simplicity, we shall neglect the mass shift throughout our calculations. The effective width of $\phi$ is then obtained as $\Gamma(s=m_\phi^2)\simeq 40~{\rm{MeV}}$.
At zero temperature, meson-loop effects vanish. Thus, we replace the self-energy of the $\phi$ meson as $\Sigma_\phi^{L,T}(s) = -im_\phi \Gamma_\phi^*(s)$.
\subsubsection{Partial restoration of chiral symmetry}
The measurements of pionic atoms strongly suggest that chiral symmetry is restored by 30\% inside the nucleus~\cite{pionicatom}.
Therefore, in the following calculations, we assume that ${f_\pi^*}^2/{f_\pi^{\rm{vac}}}^2=0.7$ holds inside the nucleus.
Given the fact that the vast majority of chiral approaches shows no significant shift of the $\phi$ meson mass in cold nuclear matter, we consider a scenario of chiral symmetry restoration in which the $f_1(1420)$ state decreases its mass toward $m_\phi$, realizing the canonical condition, ${\rm Im}\,G_V \rightarrow {\rm Im}\,G_A$.

The non-linear chiral Lagrangian based on the generalized hidden local symmetry (GHLS) describes successfully the phenomenology of the pion and vector mesons~\cite{Bando:1987br,PhysRevD.73.036001,PhysRevC.80.054912}.
The masses of vector and axial-vector states are given in the following form at leading order:
\begin{equation}
    m_A^2-m_V^2=g^2\frac{m_A^2}{m_V^2}f_\pi^2\,,
    \label{eq:mass_diff}
\end{equation}
with the gauge coupling constant $g=6.61$~\cite{Harada:2009cn}.
As noted, $m_\phi$ is assumed to stay constant. The $f_1$ mass thus carries the medium dependence responsible for the restoration of chiral symmetry. The above mass formula provides a clear relationship between the $m_{f_1}$ and the order parameter $f_\pi$.
We employ the constant-width approximation of the $f_1$ meson above threshold:
\begin{eqnarray}
\Gamma_{f_1(1420)}^{\rm vac}(s)
&=& \Theta(s - (m_\pi + 2m_K)^2)\Gamma_{f_1(1420)}^{\rm vac}\,,
\nonumber\\
\Gamma_{f_1(1420)}^{\rm vac}
&=& 54.9\,{\rm MeV}\,.
\end{eqnarray}
Following the prescription given in Ref.~\cite{SASAKI2020135172}, we shall model the in-medium width of $f_1$ as
\begin{eqnarray}
\Gamma_{f_1(1420)}^*(s)=
&&\Gamma_{f_1(1420)}^{\rm{vac}}\left(\frac{f_\pi^{*}}{f_\pi^{\rm{vac}}}\right)^2\nonumber\\
&&+\Gamma_{\phi}^*(s)\left[ 1-\left(\frac{f_\pi^{*}}{f_\pi^{\rm{vac}}}\right)^2\right]\,,
\end{eqnarray}
which ensures the degenerate widths, $\Gamma_\phi^* = \Gamma_{f_1}^*$, at the chiral restoration.

\begin{figure*}[t!]
\begin{center}
\includegraphics[width=180mm]{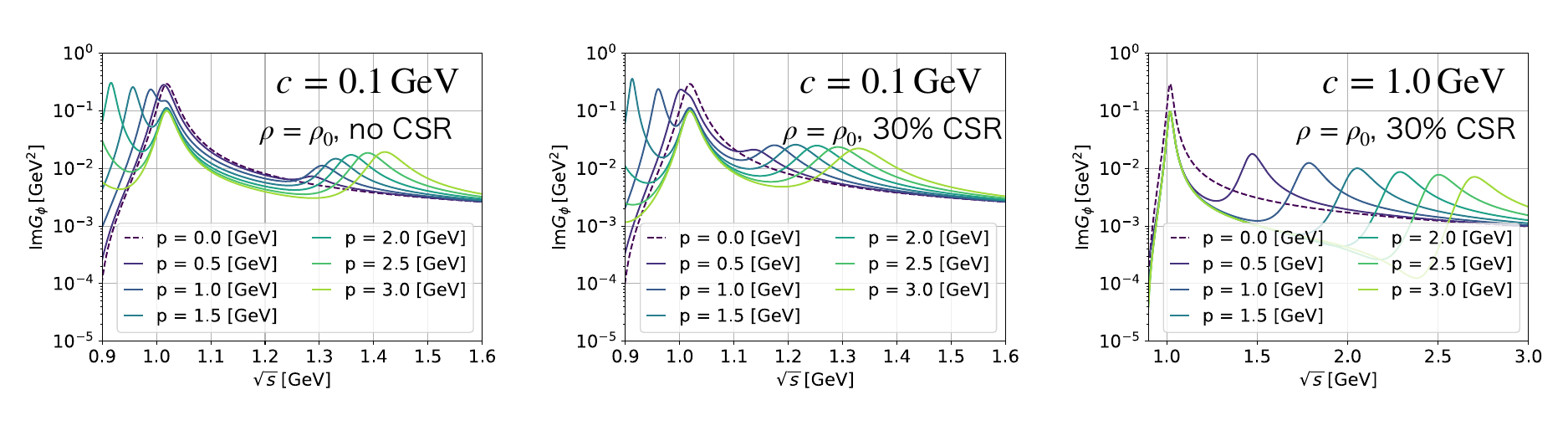}%
\caption{\label{spectral}
The spectral function of $\phi$ meson with chiral mixing at normal nuclear matter density $\rho_0$. 
Left: No chiral symmetry restoration for the mixing strength $c=0.1$ GeV is assumed, i.e. ${f_\pi^*}^2/{f_\pi^{\rm vac}}^2 = 1$. 
Center: 30\% chiral symmetry restoration for the same mixing strength is assumed, i.e. ${f_\pi^*}^2/{f_\pi^{\rm vac}}^2 = 0.7$.
Right: 30\% chiral symmetry restoration for the mixing strength of $c=1.0$ GeV is assumed. 
}
\end{center}
\end{figure*}

The full in-medium spectral function of the $\phi$ meson is presented in Fig.~\ref{spectral} for the mixing strengths $c=0.1$~GeV and $1$~GeV at $\rho=\rho_0$ with various three momenta $\vec{p}$.
The spectra exhibit the characteristic three-maxima structure at finite $\vec{p}$: the low-lying peak corresponds to the transverse component of $\phi$, the middle one to the longitudinal $\phi$ which is unmodified in our current model, and the high-lying one to the transverse component of the $f_1$ counterpart.
One readily sees a shift of the transverse vectors according to the modified dispersion relations for finite $\vec{p}$, Eq.~(\ref{eq:dispersion_rel}).
Also, it is evident with a systematic downward shift of the transverse $f_1$ contribution that the partial restoration of chiral symmetry becomes manifest in the spectra.
One should realize that for a larger mixing, $c=1$~GeV, the $f_1$ peaks are less localized, depending strongly on $\vec{p}$.

\subsection{$dN/dpd\rho dt$ distribution}
\begin{figure}[t!]
\begin{center}
\includegraphics[width=80mm]{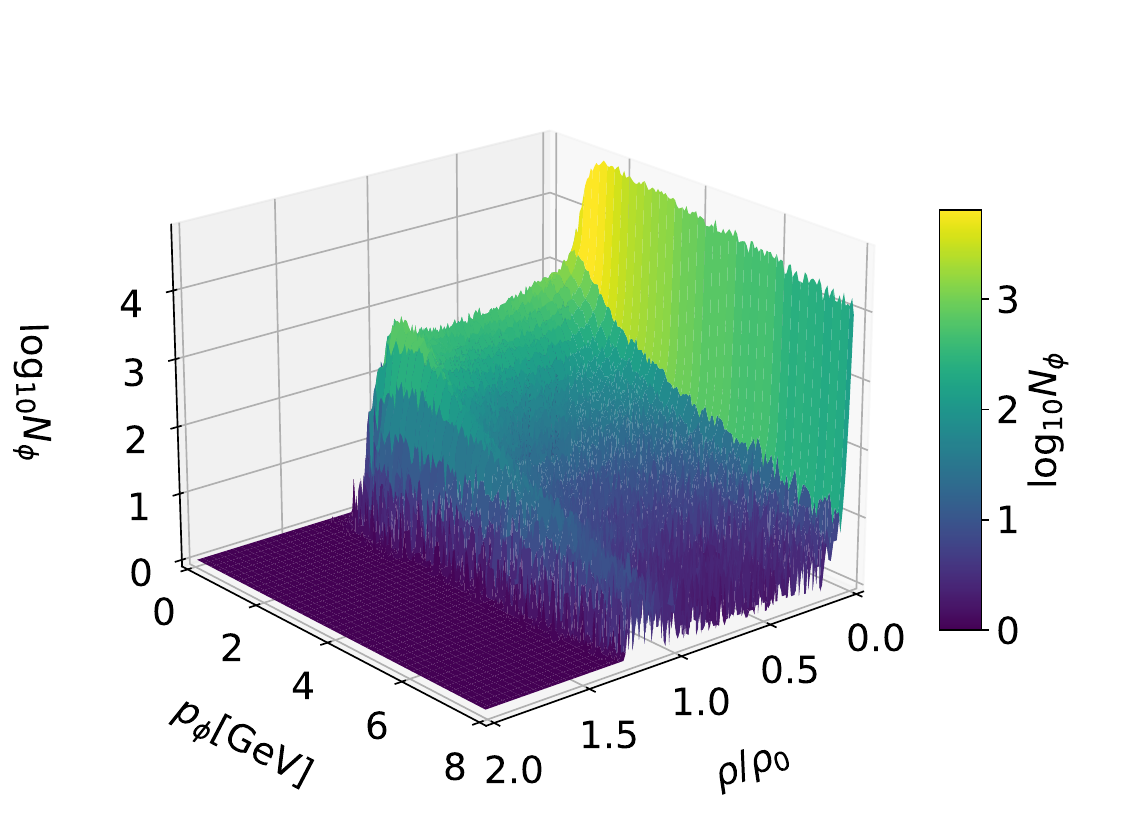}%
\caption{\label{fig:phsd}
The kinematic distribution of $\phi$ meson when it decays, calculated in the PHSD transport approach for the $30$~GeV p+Pb reactions.
}
\end{center}
\end{figure}

\begin{figure}[t!]
\begin{center}
\includegraphics[width=90mm]{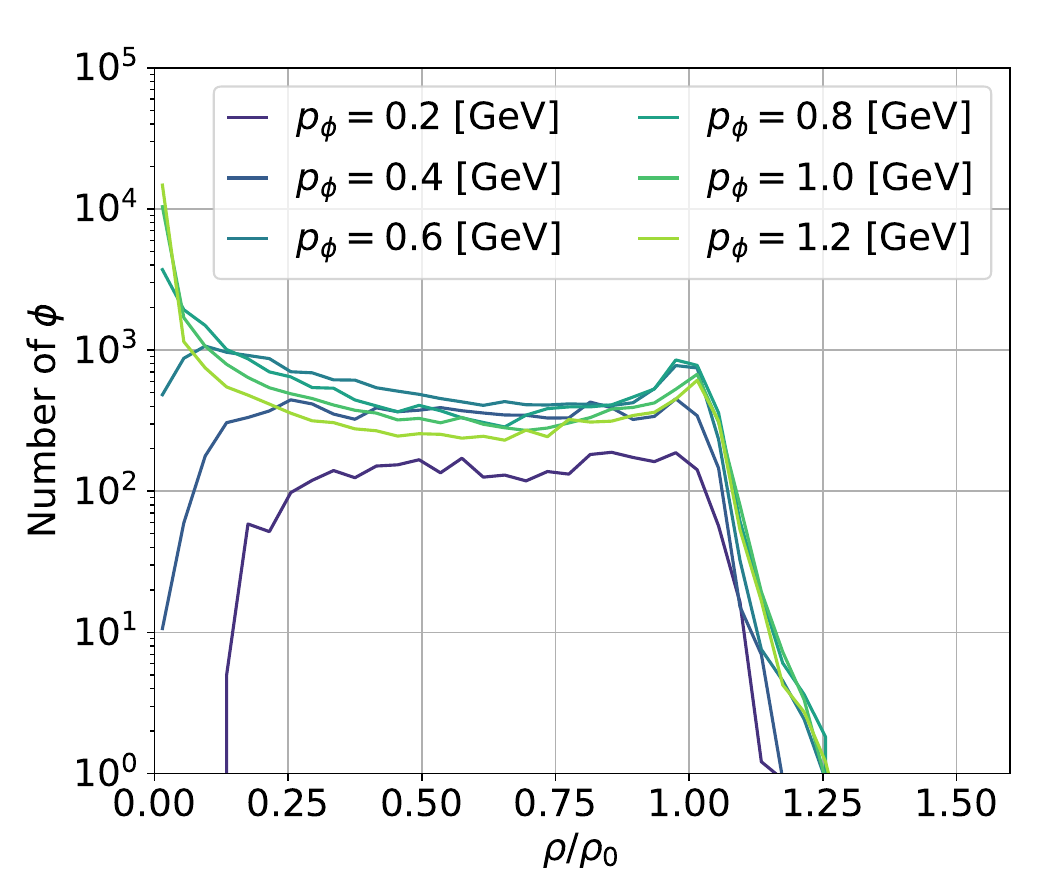}%
\caption{\label{fig:phsd_2d}
Density distribution of the decaying $\phi$ mesons for various three momenta.
}
\end{center}
\end{figure}

The $dN/dpd\rho dt$ distribution is calculated using Parton-Hadron String Dynamics (PHSD) transport approach \cite{Cassing:2008sv,Cassing:2009vt}, 
which solves the equation of motion of the $\phi$ meson (and all other particles participating in the collision) including off-shell effects to take scattering processes of the $\phi$ meson inside the nucleus into account and to calculate the momentum and density at the moment of its decay into di-electrons. 
The resulting distributions are shown in Figs.~\ref{fig:phsd} and \ref{fig:phsd_2d}. 

As can be expected from its relatively long lifetime (46.4~fm in vacuum), we can see that almost all $\phi$ mesons decays outside of the target nucleus, that is, around $\rho = 0$. 
At larger densities, the distribution is largely flat with a small peak structure at $\rho = \rho_0$, and quickly dropping to zero at even larger density. 
The fact that some $\phi$ mesons are found at densities above $\rho_0$ can be understood from their production at the initial stage of the pA collision, in which the incoming proton produces an environment with a locally slightly increased density. 

\subsection{Background and detector's response}
The background was estimated using the Monte Carlo event generator JAM~\cite{PhysRevC.61.024901} to simulate p+A 30 GeV collisions and Geant4 to simulate the interactions of the outgoing particles with the detector of the J-PARC E16 experiment. 
The background processes considered in this study are the Dalitz decay of $\pi^0$, cases of mistakenly identifying a charged pion as an electron, $\gamma$ conversion, and combinatorial background. 
The $\phi$ meson signal was scaled to match the designed yields of J-PARC E16 (15~k in Run1 and 69~k in Run2 with Cu targets~\cite{TDR2016}).
Also, the background obtained in this way was scaled to Run1/Run2 statistics, which is estimated with the charged pion rejection efficiency of the E16 spectrometer. In this calculation, we used 0.2\% as the efficiency.
Note that the spectrum of the $\phi$ meson and its chiral partner are calculated based on the chiral effective theory, while the background is estimated completely independently, and finally added as shown in Eq.~(\ref{eq:resolution_conv}). 
In reality, however, various particles interact strongly in nuclear matter and create many resonances, which could contribute to the background. 
We have confirmed that the background of the KEK-PS E325 experiment can be approximately reproduced by our simulation, and hence will assume that the obtained background is not that far from reality. 
The final convolution integral in Eq.~(\ref{eq:resolution_conv}) is performed with the mass resolution estimated for the E16 spectrometer.
Although the mass resolution in geneal depends on the momentum,
for simplicity, we assume that it is same for all momenta.
The mass resolution of 8 MeV will be employed in the following calculations as a conservative estimate with the E16 spectrometer for a $\phi$ meson carrying momenta below 1 GeV.

\section{\label{result}Result and Discussion}
In this section, we show the results of the calculations outlined in the previous Section \ref{estimation}, assuming the statistics of the co-called Run2 of the J-PARC E16 experiment, which is expected to start within a few years. 
We made this choice because only the statistics obtained in this second run of the E16 experiment are adequate to generate any potentially observable signal from the chiral mixing phenomenon, as will be shown below. 
Since the E16 experiment will probe Cu and Pb nuclei as targets (other smaller targets such as C nuclei will also be studied, but are omitted here, as they are less sensitive to density effects), the invariant mass distributions for these two cases will be discussed. 
We will also examine different values for the mixing strength $c$, ranging between those obtained from the WZW action and holographic QCD, to study the dependence of the signal on this parameter.

Before presenting the specific results, we below summarize the main features of chiral mixing at finite density and the setup of the J-PARC E16 experiment.
\begin{itemize}
 \item Chiral mixing in dense matter
   \begin{itemize}
    \item Chiral mixing and the related modified dispersion relations depend on the mixing strength $c$ and the particle 3-momentum $\vec{p}$ through the combination of $c|\vec{p}|$.
    \item Chiral mixing is enhanced when the mass difference between the $\phi$ and $f_1(1420)$ decreases.
    \item The mixing strength $c$ of the parity partners is proportional to the density, as $c = \gamma \frac{\rho}{\rho_0}$. 
    Estimates for the parameter $\gamma$ range from 0.1\,GeV (WZW action \cite{Harada:2009cn}) to 1.0\,GeV 
    (holographic QCD \cite{PhysRevLett.99.141602})\@.
  \end{itemize}
 \item J-PARC E16 experiment
  \begin{itemize}
    \item Studies fixed-target 30~GeV pA reactions with C, Cu, and Pb targets.
    \item Has a di-electron spectrometer specialized to measure the $\phi$ meson and other vector mesons in the same mass region.
    \item As the target sizes are smaller than the typical path length of the $\phi$ meson before it decays, the majority of the $\phi$ mesons produced in the pA reactions will decay outside of the target. 
    \item While $\phi$ mesons with larger momenta will more likely decay outside of the target, their mixing effect will be enhanced due to the mixing term being proportional to $|\vec{p}|$ (see above).
  \end{itemize}
\end{itemize}

\subsection{Cu target}
\begin{figure*}[t!]
\begin{center}
\includegraphics[width=180mm]{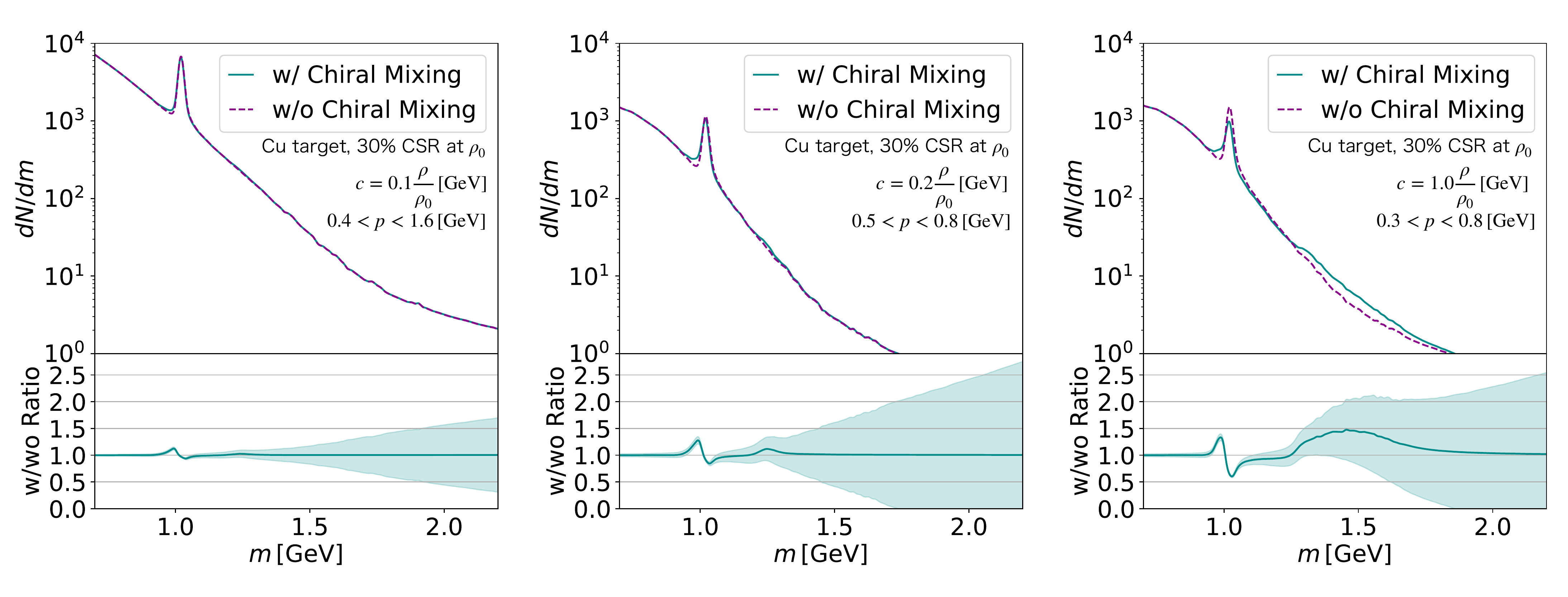}%
\caption{\label{fig:Cu}
The invariant mass spectrum for the Cu target for various mixing strengths. 
Left: The mixing strength $c=0.1 \rho/\rho_0$ GeV and the momentum range from $0.4$~GeV to $1.6$~GeV were taken. 
Center: $c=0.2 \rho/\rho_0$~GeV and the momentum range from $0.5$~GeV to $0.8$ GeV were taken. 
Right: $c=1.0 \rho/\rho_0$~GeV and the momentum range from $0.3$~GeV to $0.8$~GeV were taken. 
Each momentum range was optimally selected such that the signal of $f_1(1420)$ meson is observed with maximal significance $\sigma$.
}
\end{center}
\end{figure*}

We begin with the results for the Cu target.
The corresponding di-electron invariant mass spectra generated from 30 GeV pCu collisions are shown in Fig.\,\ref{fig:Cu}. 
In the upper half of Fig.\,\ref{fig:Cu}, the purple dotted line shows the invariant mass distribution without considering chiral mixing at finite density, while the green solid line shows the invariant mass distribution with chiral mixing. 
The lower half of Fig.\,\ref{fig:Cu} shows the ratio of the two distributions, with the statistical uncertainty indicated as cyan shaded regions.

To draw the solid lines in the plots, we assumed that the chiral symmetry is restored by 30\% at normal nuclear matter density, which brings the $f_1(1420)$ properties closer to the $\phi$ mass and width. 
In the three shown plots, three different values are taken to make a case study: $c=0.1 \rho/\rho_0$, $0.2 \rho/\rho_0$, and $1.0 \rho/\rho_0$ GeV, 
covering the range of values predicted from the WZW action and holographic QCD\@. 
As for the momentum range, as will be shown later in Fig.\,\ref{fig:heatmap}, the momentum region for which the mixing signal is 
most significant depends on the mixing strength parameter $c$. 
For this reason, the invariant mass distribution shown here and in the following sections, is calibrated to the momentum range where the mixing signal is most prominent.

For small mixing strengths ($c = $ $0.1\rho/\rho_0$~GeV and $0.2\rho/\rho_0$~GeV,
shown in left and middle plots of Fig.\,\ref{fig:Cu}) the $f_1(1420)$ signal is almost invisible. 
If we look closely at the case for $c = 0.2\rho/\rho_0$~GeV,
we can faintly see a small structure around 1.25~GeV, which, however, is not a significant enough structure to be easily confirmed experimentally. 
On the other hand, for the large mixing strength of $c = 1.0\rho/\rho_0$~GeV, the $f_1(1420)$ appears in the spectrum as a broad structure from 1.25~GeV to 1.75~GeV\@. 
However, even though the ratio of the distributions with/without chiral mixing is large, the corresponding statistical uncertainty turns out to be sizable as well, such that the $f_1(1420)$ is only visible at about 1$\sigma$ significance. 
Furthermore, due to the strongly modified dispersion relation (see Eq.\,(\ref{eq:dispersion_rel})), the new structure appearing in the spectrum that corresponds to the $f_1(1420)$, is broad and distributed around its vacuum mass. It is therefore in such a measurement not possible to discuss the degeneracy of the chiral partner associated with the restoration of chiral symmetry. 

One could here envisage several strategies to improve this situation. 
Choosing for example a narrower momentum range at smaller momentum values, one can in principle expect to see a peak-like structure below 1.42 GeV, similar to the spectral function shown in Fig.\,\ref{spectral}. 
This, however, will increase the statistical uncertainty. 
Furthermore, since chiral mixing always appears in the form of $c\vec{p}$, the signal will be suppressed in the small absolute momentum region. 
Conversely, selecting a larger absolute momentum region will enhance the signal due to the larger mixing strength, but at the same time also suppress it, as a larger fraction of $\phi$ mesons will decay outside of the target nucleus. 
All this leads to an intermediate momentum range, for which the sensitivity to the mixing signal is most enhanced, which was chosen to generate the results shown in Fig.\,\ref{fig:Cu}.

It is moreover interesting to observe that due to the modified dispersion relation of the transverse $\phi$ meson mode, which effectively reduces its mass, 
a tail on the left side of the peak is generated, while the longitudinal $\phi$ meson mode peaks at the vacuum mass value of 1.02~GeV\@.
Previously, such a tail has been considered to be caused by a $\phi$ meson mass shift, possibly related to the partial restoration of chiral symmetry in nuclear matter \cite{PhysRevD.90.094002}. Our results show that the chiral mixing effect at finite density can also produce such a tail. 
As in the chiral mixing scenario considered here only the transverse mode generates this tail, it is in principle possible to distinguish the two cases by looking at the angular distribution of the outgoing dileptons (see Ref.\,\cite{Park:2022ayr} for a more detailed discussion).

The Cu target results are summarized below. 
\begin{itemize}
 \item Although the $f_1(1420)$ signal is 
 rather significant in the spectral function at finite density (see Fig.\,\ref{spectral}), it will be difficult to observe it for the small mixing strength cases.
 The main reasons for this difficulty are the long $\phi$ meson lifetime and the relatively small size of the Cu target, which cause a large fraction of the produced $\phi$ mesons to decay outside of the dense target. 
 They will hence not be very sensitive to density effects. One obvious strategy to improve this situation would be to choose a larger target, one of which will be discussed in the following section.

 \item The large mixing strength makes a clear difference, by generating a visible, but rather broad structure in the di-lepton spectrum. 
 The broadness of this structure can be understood from the significantly modified dispersion relation, which makes the $f_1(1420)$ peak position strongly momentum dependent, as can be seen in the right plot of Fig.\,\ref{spectral}. 
 This unfortunately makes it difficult to study the movement of the $f_1(1420)$ mass towards the mass degeneracy with its chiral partner, because the $f_1(1420)$ signal is spread approximately symmetrically around its vacuum mass.

  \item Chiral mixing also leaves an effect on the small mass side of the $\phi$ meson invariant mass distribution due the modified dispersion relation. 
  Thus, care has to be taken when interpreting a tail in the $\phi$ meson spectrum as it could be a result of both a genuine $\phi$ meson mass shift or chiral mixing (or a mixture of both). 
  It is, however, in principle possible to distinguish the two cases by either studying the momentum dependence of the tail or by disentangling the longitudinal and transverse polarization modes.
\end{itemize}

\subsection{Pb target}
\begin{figure*}[t!]
\begin{center}
\includegraphics[width=180mm]{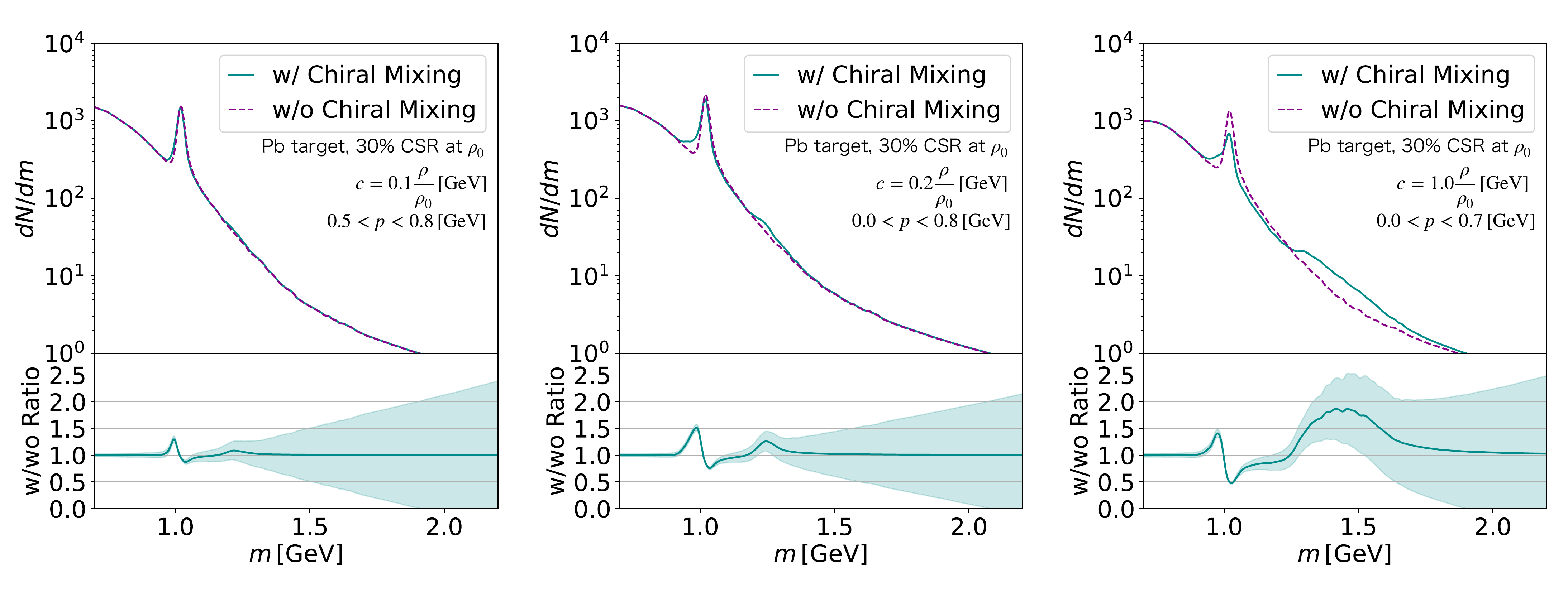}%
\caption{\label{fig:Pb}
The same as in Fig.~\ref{fig:Cu} but for the Pb target.
}
\end{center}
\end{figure*}

To possibly enhance the sensitivity to the $f_1(1420)$ mixing signal, we will next study the Pb target. 
With its larger nuclear radius, the probability to decay inside the nucleus will increase, which should enhance the mixing signal. 
However, as we have already confirmed for the Cu target case, with a wide momentum range and a large mixing strength, the $f_1(1420)$ distribution broadens, which will make it difficult to discuss the degeneracy of the chiral partners. 
The results for the Pb target are shown in Fig.\,\ref{fig:Pb}.

The statistics assumed in this calculation are for the Pb target with the same beamtime as for the Cu target planned for the J-PARC E16 experiment Run2.

Compared to the Cu case, the larger Pb target increases the probability of the $\phi$ meson to decay inside the target nucleus. 
The $f_1(1420)$ signal is therefore somewhat easier to see for all mixing strengths. The result for the $c=0.1 \rho/\rho_0$~GeV case (left plot in Fig.\,\ref{fig:Pb}) does not show any significant change, but for $c=0.2 \rho/\rho_0$~GeV (middle plot), a clearly better signal is observed. 
Furthermore, since the dispersion relation does not change much for this mixing strength, the $f_1(1420)$ structure for different momenta remains approximately in the same mass region, specifically at about 1.2~GeV, which is closer to the $\phi$ meson mass than in vacuum. 
The chiral symmetry restoration and mass degeneracy calculations were performed  
as shown in Section \ref{estimation},
assuming the chiral symmetry restoration by 30\% at normal nuclear matter density. 
If this scenario is realized in nature, with a mixing strength only slightly larger than expected by the WZW action, our result suggests that with enough statistics to make the $f_1(1420)$ signal significant in this experimental setup, it would be possible to discuss and experimentally study the phenomenon of degenerating chiral partners.
The magnitude of the statistical uncertainty in the calculation indicates that the $f_1(1420)$ is observable with about a significance of 2$\sigma$. Our result therefore provides motivation for a large statistics measurement for the Pb target at the J-PARC E16 experiment and to look for the same mixing signal at other experimental facilities in the future.

In the $c=1.0 \rho/\rho_0$~GeV case, it remains difficult to discuss the chiral partner degeneracy because of the broadness of the structure corresponding to the $f_1(1420)$. 
However, even in this case, since a larger fraction of $\phi$ mesons decay inside the nucleus, the $f_1(1420)$ signal is enhanced and can be seen with a higher significance compared to the Cu target.

Let us here summarize the results for the Pb target. 
\begin{itemize}
 \item Even with a relatively small mixing strength of $c=0.2 \rho/\rho_0$~GeV, the larger medium size makes it easier to observe the $f_1(1420)$ signal (see middle plot of Fig.\,\ref{fig:Pb}). 
 This happens also because the dispersion relation does not change too much with this mixing strength, hence the $f_1(1420)$ mass does not strongly depend on the momentum. 
 We thus find that the $f_1(1420)$ signal appears around 1.25~GeV,
 assuming the restoration of chiral symmetry by 30\% at normal nuclear matter density, where the two chiral partners are starting to approach each other.
    
 \item For the large mixing strength of $c=1.0 \rho/\rho_0$~GeV, the $f_1(1420)$ signal is enhanced compared to the Cu target, but is still too broad to carry meaningful information regarding the chiral partner degeneracy at finite density.
\end{itemize}

\subsection{Mixing strength dependence of \\ the significance of the $f_1(1420)$ signal} 
\begin{figure}[t!]
\includegraphics[width=90mm]{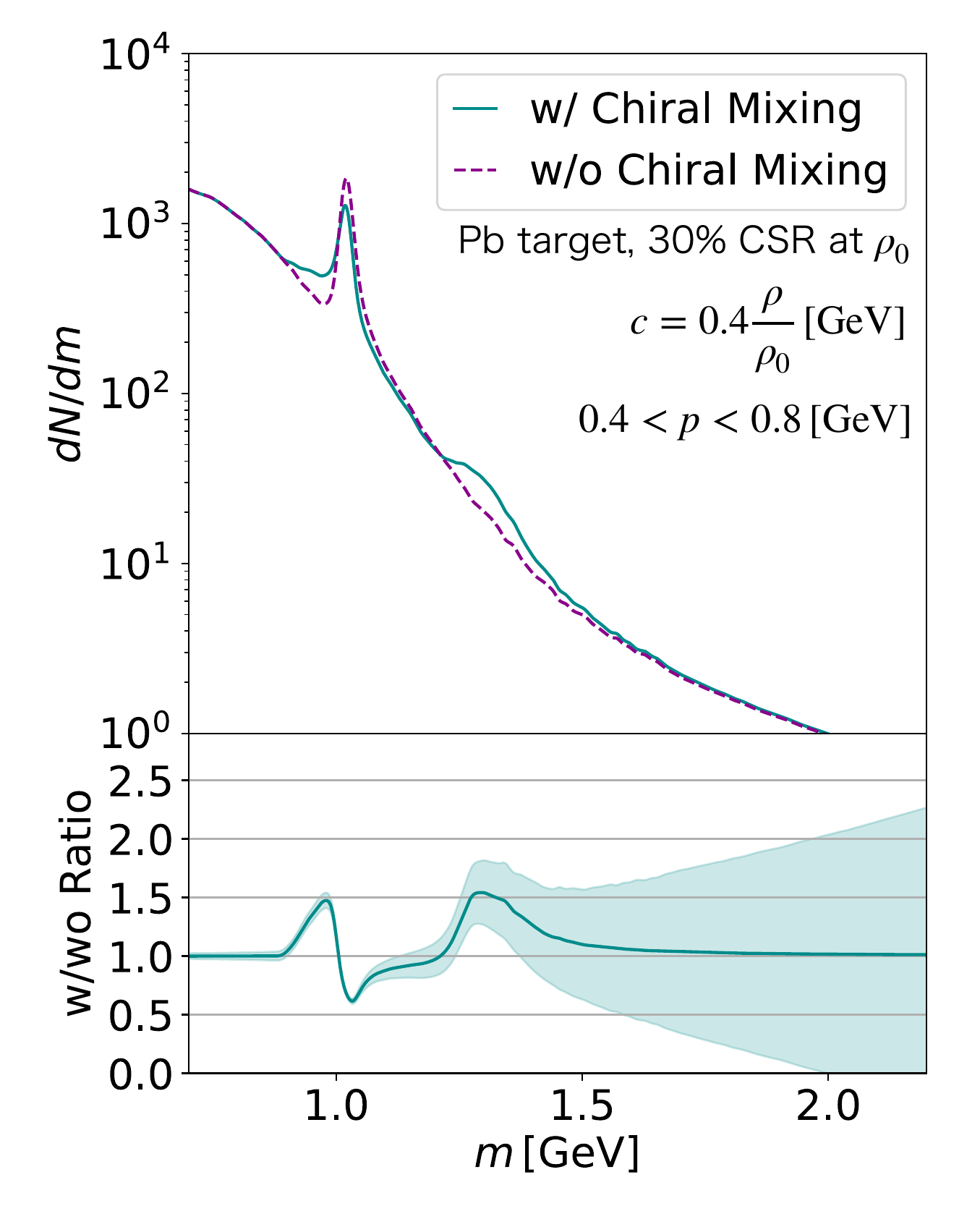}
\caption{\label{fig:c04} 
The same as in Fig.~\ref{fig:Pb} but for $c=0.4 \rho/\rho_0$ GeV.
}
\end{figure}

It is currently not known what the exact value of the mixing parameter $c$ is. As we discussed in Section \ref{theory}, the value predicted by holographic QCD ($c=1.0 \rho/\rho_0$ GeV) may be artificially enhanced by the large $N_c$ approximation, while we expect the value derived by the WZW action in the mean field approximation ($c=0.1 \rho/\rho_0$ GeV) to be more realistic.
However, since there is no reliable QCD proper constraint on $c$ at finite density, we have in this work tested the whole range between $0.1\rho/\rho_0$ GeV and $1.0\rho/\rho_0$ GeV. 
As a result, we found that the case with $c=0.4 \rho/\rho_0$ GeV led to a $f_1(1420)$ signal with the highest confidence level. 
The dilepton spectrum for this case is shown in Fig.\,\ref{fig:c04}. 

This result shows again that a stronger mixing strength does not necessarily enhance the $f_1(1420)$ signal. 
To recapitulate, this outcome can be understood from a competition of two effects of the mixing term in Eq.\,(\ref{eq:mix}) and its coefficient $c$. Increasing $c$ leads to a stronger mixing effect, which enhances the $f_1(1420)$ signal. 
On the other hand, a larger $c$ value also induces a strong modification of the dispersion relations of the related particles, and hence to a momentum dependent mass [see Eq.\,(\ref{eq:dispersion_rel})],
which causes the signal to be distributed over a larger mass range due to the momentum intergral in Eq.\,(\ref{eq:InvMass}).
The combination of these two effects leads to a signal that is most significant for an intermediate value of $c$. 
The situation is in fact further complicated by our assumption of the chiral symmetry being restored by 30\% at normal nuclear matter density, which makes the $f_1(1420)$ mass density dependent [see Eq.(\ref{eq:mass_diff})]. 
As can be seen in Figs.\,\ref{fig:phsd} and \ref{fig:phsd_2d}, different momenta lead to different density distributions and hence not only to a modified value of $c$, but also to an $f_1(1420)$ mass that is closer to the one of the $\phi$ meson, its chiral partner. 
Therefore, choosing an ideal momentum range for the measured dilepton spectrum is crucial for obtaining significant mixing signal.
We found that, for an absolute momentum range of about $0.6\pm0.2$~GeV,
the modified dispersion relation and the density dependent $f_1(1420)$ mass conspire to generate the strongest signal for the mixing strength of $c=0.4 \rho/\rho_0$~GeV, as shown in Fig.\,\ref{fig:c04}.

\begin{figure*}[t!]
\begin{center}
\includegraphics[width=180mm]{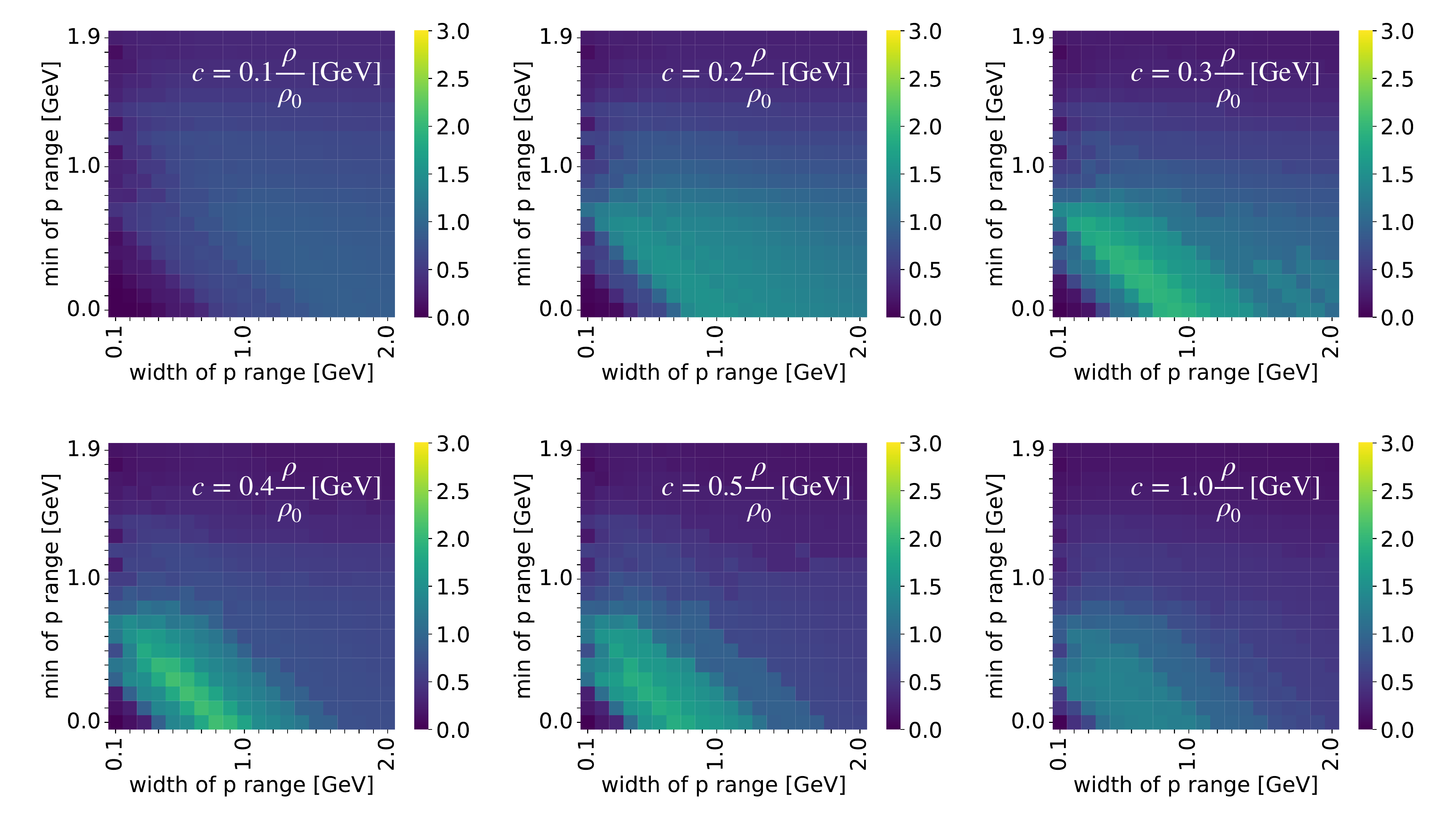}%
\caption{\label{fig:heatmap}
The heatmap of the statistical significance observing the $f_1(1420)$ meson for various mixing strengths with the Pb target under the E16 Run2 statistics.
}
\end{center}
\end{figure*}

The statistical significance of the mixing signal is illustrated more generally for different momentum ranges and mixing strengths in 
Fig.\,\ref{fig:heatmap}. 
In this figure, the vertical axes represent the lower limit of the momentum range of the invariant mass distribution, and the horizontal axes its width. 
As can be observed from the different plots, the $f_1(1420)$ signal is strongest for a mixing strength of about $c=(0.4\pm0.1)\rho/\rho_0$~GeV\@. 
In addition, it is understood that a proper choice of both absolute momentum value and range width is crucial to obtain a strong signal, which, with the statistics assumed in this work, can be as large as 2.5$\sigma$.

Furthermore, the parameter areas that generate the most significant signals form a band which has a slope of approximately -1. For this band, the sum of the vertical and horizontal axes, which is nothing but the upper limit of the momentum range, is constant, with a value of about 1~GeV\@. 
This means that if a certain momentum is set as the upper limit of the momentum range, the significance will remain the same even if the lower limit changes to some extent, which can be interpreted as follows. 
The absolute momentum around 1 GeV is most sensitive to the mixing effect, as with this momentum a sufficient fraction of the vector meson and its chiral partner can decay inside the nuclear target, and at the same time enough statistics can be accumulated. 
If the upper limit is raised above that level, the percentage of decay outside the nucleus will increase, leading to a suppressed signal. 
Keeping the upper limit fixed, while the lower limit is brought closer to the upper one, the significance falls because of the decreased statistics, as shown in the upper left side of the diagonal structures in Fig.\,\ref{fig:heatmap} for 
$c=0.4\rho/\rho_0$~GeV and $0.5\rho/\rho_0$~GeV, for example.

Let us here briefly discuss the specific mixing strength derived from the WZW action, that is $c=0.1 \rho/\rho_0$~GeV\@. 
In this case, the $f_1(1420)$ signal is quite small and at most reaches a significance of only about 1$\sigma$. 
The corresponding invariant mass distribution shows an almost flat structure, and hence requires a precise background estimation. 
In such an estimation, we can however not dismiss the possibility that such a structure has an unknown correlated background. 
Therefore, it will be important to prepare a baseline for comparison with the experimental results, without considering chiral mixing, which is depicted by the purple short-dashed lines in the figures of the invariant mass spectra. 

In heavy-ion collision experiments, it is common to discuss the existence of a possible quark-gluon plasma phase by preparing pp collisions as a baseline in comparison with the A+A collisions. 
Therefore, it may be necessary to measure and analyze similar pp collisions even for the fixed target experiments studied here. 
Appropriately scaled, such a collision could for example be used in comparison with the results of the Pb target studied in this section, to identify the genuine density effects, including chiral mixing.

If such a precise analysis can be carried out, the present work indicates that if the mixing effect is strong enough, an axial-vector meson can be observed with a significance of about 1.0 to 2.5$\sigma$ at the Run2 of the J-PARC E16 experiment. 
Furthermore, if an axial-vector meson is observed and chiral symmetry is restored by 30\,\% at normal nuclear matter density, our results suggest that the mass change is large enough to experimentally study the approach toward the degeneracy 
of the vector and axial-vector chiral partners.

The results of this section are summarized below. 
\begin{itemize}
 \item For both too small and too large mixing strengths, it will be difficult to observe the $f_1(1420)$ signal in the dilepton spectrum.
 \item For an intermediate mixing strength of $c=(0.4\pm0.1)\rho/\rho_0$~GeV, the mass of the $f_1(1420)$ signal in different momentum regions happens to coincide, which can be understood from a cancellation of effects of the modified dispersion relation and the movement towards degeneracy of the chiral partners. This enhances the significance of the chiral mixing signal, which (if such a mixing strength is realized in nature) hence may be observed in a high-statistics measurement at J-PARC\@. 
 \item An ideal choice of the momentum range included in the dilepton spectrum analysis is crucial for obtaining a significant 
 mixing signal. 
 For the specific 30 GeV pA collisions studied in this work, the contribution from the (lab frame) momentum region around and below 1~GeV is indispensable for generating an observable signal.
 \item With an appropriate momentum range and an ideal mixing strength, the $f_1(1420)$ can be observed with a statistical significance of about 2.5$\sigma$ with J-PARC E16 Run2 statistics and a Pb target. 
 In such a case, the $f_1(1420)$ structure is furthermore narrow enough to study its (movement towards) degeneracy with its chiral partner, the $\phi$ meson.
\end{itemize}

\subsection{Partial restoration of chiral symmetry in the $\phi$ sector}
\begin{figure}[t!]
\includegraphics[width=90mm]{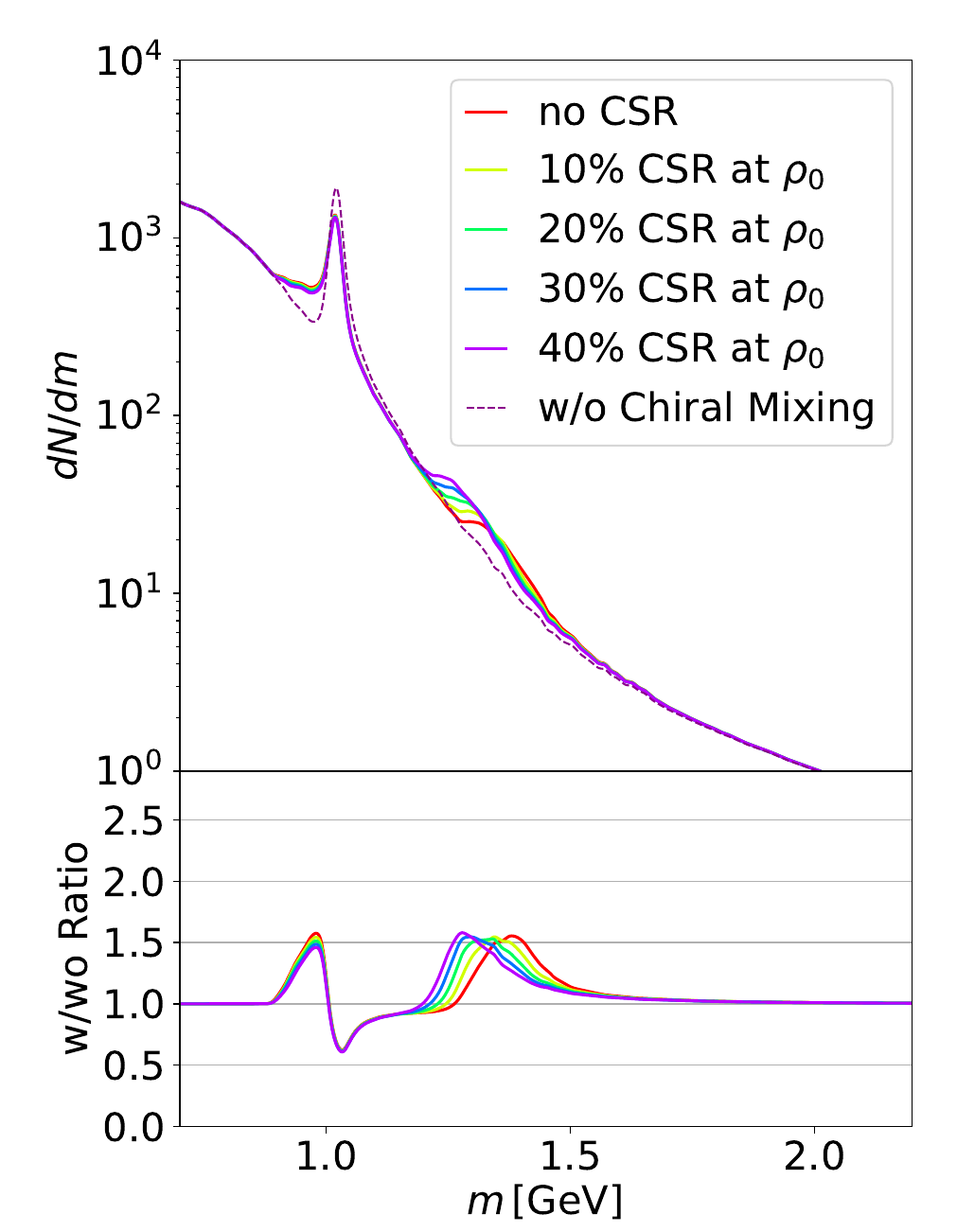}
\caption{\label{fig:c04_var} 
The invariant mass spectrum for various degrees of chiral symmetry restoration with the mixing strength of $c=0.4 \rho_0/\rho$ GeV.}
\end{figure}

In the previous calculations, we have assumed that the chiral symmetry is restored by 30\% at normal nuclear matter density $\rho_0$. It is plausible that the restoration could be somewhat delayed in the strange-quark sector compared to that of the u- and d-quarks. 
Here, we compute the di-lepton invariant mass spectrum by reducing the degree of chiral symmetry restoration. 
As shown in Fig.~\ref{fig:c04_var}, 
we find that, even with a 10\% restoration, the contribution from the $f_1(1420)$ meson moves closer toward the $\phi$ peak. 
Since the E16 spectrometer has a mass resolution of a few MeV (the exact value is momentum dependent), 
our results demonstrate the feasibility to discriminate between the vacuum $f_1(1420)$ and the modified one according to Eq.~(\ref{eq:dispersion_rel}).

\section{\label{summary}Summary}
In this paper, we proposed a new method to experimentally verify the relationship between spontaneous chiral symmetry breaking and hadron masses by observing the chiral partners $f_1(1420)$ and $\phi$ through a finite-density-induced chiral mixing mechanism \cite{SASAKI2020135172,PhysRevD.106.054034}. 
As a primary example, we calculated the contribution of this chiral mixing to the expected invariant mass distribution generated from 30~GeV pA collisions to be probed at the J-PARC E16 experiment.
As a result, we found that depending on the mixing strength, 
the $f_1(1420)$ signal may be observable in the dilepton spectrum with a statistical significance of about 1 to 2.5$\sigma$ with the statistics achievable at the J-PARC E16 experiment Run2. 
Furthermore, assuming that the chiral symmetry is restored by 30\,\% at normal nuclear matter density, and that the $f_1(1420)$ and the $\phi$ meson are chiral partners,
we observed that the effect of the $f_1(1420)$ mass decreasing towards the $\phi$ meson is large enough to be measurable if the mixing signal is seen. 
Such a measurement would constitute an important step towards a deeper understanding of the relation between hadron masses and the spontaneously broken chiral symmetry.

\begin{acknowledgments}
\end{acknowledgments}
We thank all the J-PARC E16 collaborators, especially the spokesperson Satoshi Yokkaichi, for useful discussions and giving us the environment of Monte Carlo simulations under the E16 conditions. 
We also thank Elena Bratkovskaya for fruitful discussions related to the use of the PHSD transport approach.
This work was supported by Grant-in-Aid for JSPS Fellows Grant No. JP23KJ1650. This work is also supported in part by the WPI program “Sustainability with Knotted Chiral Meta Matter (WPI-SKCM$^2$)” at Hiroshima University.
The work of C.S. was supported partly by the Polish National Science Centre (NCN) under OPUS Grant No. 2022/45/B/ST2/01527. 
P.G. is supported by the Grant-in-Aid for Scientific Research (A) (JSPS KAKENHI Grant Number JP22H00122).
K.S. is supported by the Grant-in-Aid for Scientific Research on Innovative Areas and the Grant-in-Aid for Scientific Research (A) (JSPS KAKENHI Grant Numbers JP18H05401 and JP20H00163).

\nocite{*}

\bibliography{mybibfile}

\end{document}